\documentstyle[preprint,aps,epsf]{revtex}
\tightenlines
\begin{document}
\newcommand{\be}{\begin{equation}}
\newcommand{\ee}{\end{equation}}
\newcommand{\bea}{\begin{eqnarray}}
\newcommand{\eea}{\end{eqnarray}}
\title{Partition function zeros of the $Q$-state Potts model 
for non-integer $Q$}
\author{Seung-Yeon Kim\footnote{E-mail: kim@cosmos.psc.sc.edu} 
and Richard J. Creswick\footnote{Corresponding author. 
E-mail: creswick.rj@sc.edu}}
\address{Department of Physics and Astronomy,
University of South Carolina,\\ Columbia, South Carolina 29208, USA}
\author{Chi-Ning Chen and Chin-Kun Hu}
\address{Institute of Physics, Academia Sinica, Nankang, Taipei,
Taiwan 11529, Republic of China} 
\maketitle
\begin{abstract}
The distribution of the zeros of the partition function in the complex 
temperature plane (Fisher zeros) of the two-dimensional $Q$-state Potts 
model is studied for non-integer $Q$.
On $L\times L$ self-dual lattices studied ($L\le8$), no Fisher 
zero lies on the unit circle $p_0=e^{i\theta}$ in the complex 
$p=(e^{\beta J}-1)/\sqrt{Q}$
plane for $Q<1$, while some of the Fisher zeros lie on the unit circle
for $Q>1$ and the number of such zeros increases with increasing $Q$. 
The ferromagnetic and antiferromagnetic properties of the Potts model 
are investigated using the distribution of the Fisher zeros. 
For the Potts ferromagnet we verify the den Nijs formula for the thermal 
exponent $y_t$.
For the Potts antiferromagnet we also verify the Baxter conjecture for the 
critical temperature and present new results for the thermal exponents
in the range $0<Q<3$.
\end{abstract}
\draft


\section{introduction}

The $Q$-state Potts model\cite{wu} on a lattice $G$ for integer $Q$ 
is defined by the Hamiltonian
\be
{\cal H}=-J\sum_{\langle i,j\rangle} \delta(\sigma_i,\sigma_j),
\ee
where $J$ is the coupling constant, 
$\langle i,j\rangle$ indicates a sum over nearest-neighbor pairs,
and $\sigma_i=1,...,Q$.
Fortuin and Kasteleyn\cite{fortuin} have shown that 
the partition function can be written as 
\be
Z=\sum_{G^\prime\subseteq G} Q^{n(G^\prime)} (e^{\beta J}-1)^{b(G^\prime)},
\ee
where the summation is taken over all subgraphs $G^\prime \subseteq G$,
and $n(G^\prime)$ and $b(G^\prime)$ are, respectively, the number of clusters
and occupied bonds in $G^\prime$.
In Eq. (2) $Q$ need not be an integer and Eq. (2) defines 
the partition function of the $Q$-state Potts model for non-integer $Q$.
In this paper we discuss the ferromagnetic (FM) and antiferromagnetic (AF)
properties of the two-dimensional Potts model for
non-integer $Q$ using the partition function zeros in the complex
temperature plane (Fisher zeros).

We have calculated the exact partition function of the 
Potts model on finite lattices of size $3\le L\le8$ by the Chen-Hu 
algorithm\cite{chen1} based on the Fortuin-Kasteleyn representation.
We have used self-dual boundary conditions\cite{chen2}
for which the Fisher zeros of the Potts model map into themselves
under the duality transformation, $p\to1/p$, where 
$p=(e^{\beta J}-1)/\sqrt{Q}$.
The self-dual lattices considered in this paper are periodic in the
horizontal direction and there is another site above the $L\times L$
square lattice, which connects to $L$ sites on the first row.  

Fig. 1 shows the Fisher zeros in the complex $p$ plane of the Potts model 
for several values of non-integer $Q$. For $Q<1$ no zero lies on the unit 
circle $p_0=e^{i\theta}$ and as $Q$ approaches one from below some zeros 
approach the unit circle.
In Fig. 1a we have omitted half of the Fisher zeros which give
no more information because they can be obtained using the duality
transformation from the Fisher zeros in the figure.
For $Q>1$ zeros begin to lie on the unit circle $p_0$ and
the number of zeros on the unit circle increases with increasing $Q$.
Fig. 1b shows the Fisher zeros in the range $1<Q<2$.
The two circles shown are the FM and AF
loci for the 2-state (Ising) model in the thermodynamic limit\cite{fisher}.
For $Q>2$ the zeros in the AF region (${\rm Re}(p)<0$) become more scattered.  
Finally, as shown in Fig. 1d, for large $Q$ all the zeros lie on the 
unit circle $p_0$\cite{chen2}.
The distribution of zeros varies continuously with $Q$ with the exception
of $Q=1$ and $Q=2$.
For $Q=1$ the zeros are all degenerate at $p=-1$\cite{chen2},
while for $Q=2$ the symmetry between the FM and 
AFM Ising model forces the AF zeros
to lie on the locus $p=-\sqrt{2}+e^{i\theta}$.


\section{the ferromagnetic Potts model}

There is no rigorous proof that the FM critical point $p_c=1$ 
for $Q<4$ except for $Q=2$\cite{wu}.
For $1<Q<4$ we observe that the zero closest to the positive real axis 
always lies on the unit circle and approaches the critical point $p_c=1$ 
in the thermodynamic limit.
For $Q<1$, however, because no zero lies on the unit circle for a
finite-size lattice, we need to calculate the FM critical point
from the zero closest to the point $p_c=1$ (the first zero).
By using the Bulirsch-Stoer (BST) algorithm\cite{bst}
we extrapolated our results for finite lattices to infinite size.
The error estimates are twice the difference between the ($n-1,1$) and
($n-1,2$) approximants.
We find that the first zero converges on the critical point $p_c=1$. 

From the first zero, $p_1$, we have calculated the thermal exponent 
$y_t(L)$ defined as\cite{creswick}
\be
y_t(L)=-{{\rm ln}\{ {\rm Abs}[p_1(L+1)-1]/{\rm Abs}[p_1(L)-1]\}
\over{\rm ln}[(L+1)/L]}
\ee 
or
\be
y_t(L)=-{{\rm ln}\{ {\rm Im}[p_1(L+1)]/{\rm Im}[p_1(L)]\}
\over{\rm ln}[(L+1)/L]}.
\ee
For $Q<1$ the imaginary part of the first zero is not a monotonic
function of $L$, and so in this range we used Eq. (3) to calculate
$y_t$. For $Q>1$ Eq. (4) was found to give the best estimate.
Fig. 2 shows the BST estimates of the thermal exponent which are in excellent 
agreement with the den Nijs formula $y_t=(3-3x)/(2-x)$\cite{dennijs}, 
where $x=(2/\pi){\rm cos}^{-1}(\sqrt{Q}/2)$.
Bl\"{o}te {\it et al.}\cite{blote} have obtained results similar 
to Fig. 2 using heat capacity data on infinitely long strips.


\section{the antiferromagnetic Potts model}

For AF interaction $J<0$ the physical interval is $0\le e^{\beta J}\le1$, 
which corresponds to $-1/\sqrt{Q}\le p\le0$.
Baxter\cite{baxter} has conjectured the existence of an AF 
critical point for $Q\le3$ at $p_c=(\sqrt{4-Q}-2)/\sqrt{Q}$, 
and we expect that in the thermodynamic
limit the locus of the Fisher zeros cuts the negative real axis 
between $-1/\sqrt{Q}$ and 0. 
For $Q=2.9$ and $L=8$ (Fig. 1c) two zeros already lie on the negative real
axis, but they are outside the physical interval. 
Table I shows the real and imaginary parts of the zero $p_a(L)$ closest
to the AF interval of the negative real axis
for $Q=0.1$ and 2.9 for even-size lattices,
and in Table I the last row are the BST estimates for $L\to\infty$.
Fig. 3 shows the BST estimates of the critical points
of the Potts antiferromagnet from the Fisher zeros for $0<Q<3$
which are in excellent agreement with the Baxter conjecture\cite{baxter}.

From the Fisher zeros we have also calculated the thermal exponent
$y_t(L)$ defined as\cite{creswick}
\be   
y_t(L)=-{{\rm ln}\{ {\rm Abs}[p_a(L+2)-p_c]/{\rm Abs}[p_a(L)-p_c]\}
\over{\rm ln}[(L+2)/L]}.
\ee
Fig. 4 shows the thermal exponent $y_t(L)$ for $0<Q<3$.
In Fig. 4 as $Q$ increases the thermal exponent decreases
and around $Q=2$ it is consistent with the known exact value $y_t=1$
of the Ising model. Note that these data are calculated from
lattices of size $L=4$, 6, and 8 only, and no attempt is made
to extrapolate to infinite size.

To our knowledge these are the first such calculations of $y_t$ 
for non-integer $Q$. Of course for the Ising model, $Q=2$,
the exact result $y_t=1$ has been known for many years, and 
our results are consistent with this. The value of $y_t$ for $Q=3$
has been the subject of some debate recently. 
Ferreira and Sokal\cite{sokal} expect $y_t=1/2$, 
while Wang {\it et al.}\cite{wang} found $y_t=0.77$.
Although it might appear from Fig. 4 that our results for $Q=3$
agree with the result of Wang {\it et al.}, we must emphasize that
these results are for very small lattices and have not been 
extrapolated to infinite size. If we include calculations for $Q=3$
using the microcanonical transfer matrix\cite{creswick} for $L=10$, 
and apply the BST algorithm our results are inconclusive.
We are currently extending our calculations to larger lattices
in order to address this interesting question.

\begin{center}
ACKNOWLEDGMENTS
\end{center}
This work was supported in part by the National Science Council
of the Republic of China (Taiwan) under grant number
NSC 88-2112-M-001-011.



\begin{table}
\caption{The real part Re($p_a$) and the imaginary part Im($p_a$)
of the zero $p_a(L)$ closest to the antiferromagnetic interval
for $Q=0.1$ and 2.9. 
The last row is the BST extrapolation to infinite size.}
\begin{tabular}{ccccc}
$Q$ &0.1 & &2.9 & \\
$L$ &Re($p_a$) &Im($p_a$) &Re($p_a$) &Im($p_a$) \\
\hline
4 &$-$0.0902065 &0.00855442  &$-$0.501335 &0.198694 \\
6 &$-$0.0844057 &0.00443054  &$-$0.507772 &0.137197 \\
8 &$-$0.0823869 &0.00272837  &$-$0.511662 &0.107013 \\
$\infty$ &$-0.078(4)$ &$-0.001(2)$ &$-0.527(5)$ &0.019(41) \\
\hline
Baxter's conjecture &$-$0.07956 &0 &$-$0.5586 &0 \\
\end{tabular}
\end{table} 


\begin{figure}
\epsfbox{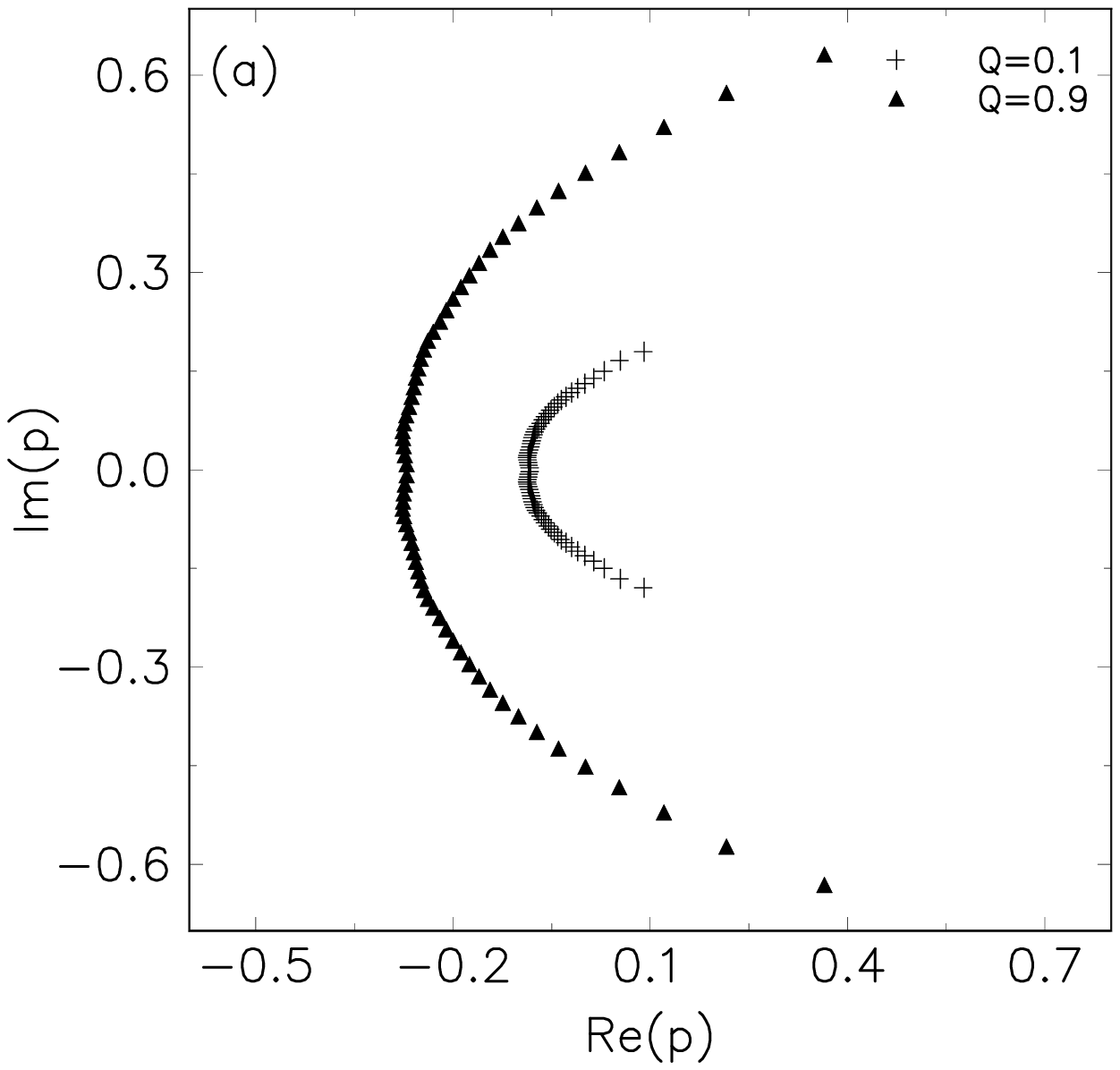}
\epsfbox{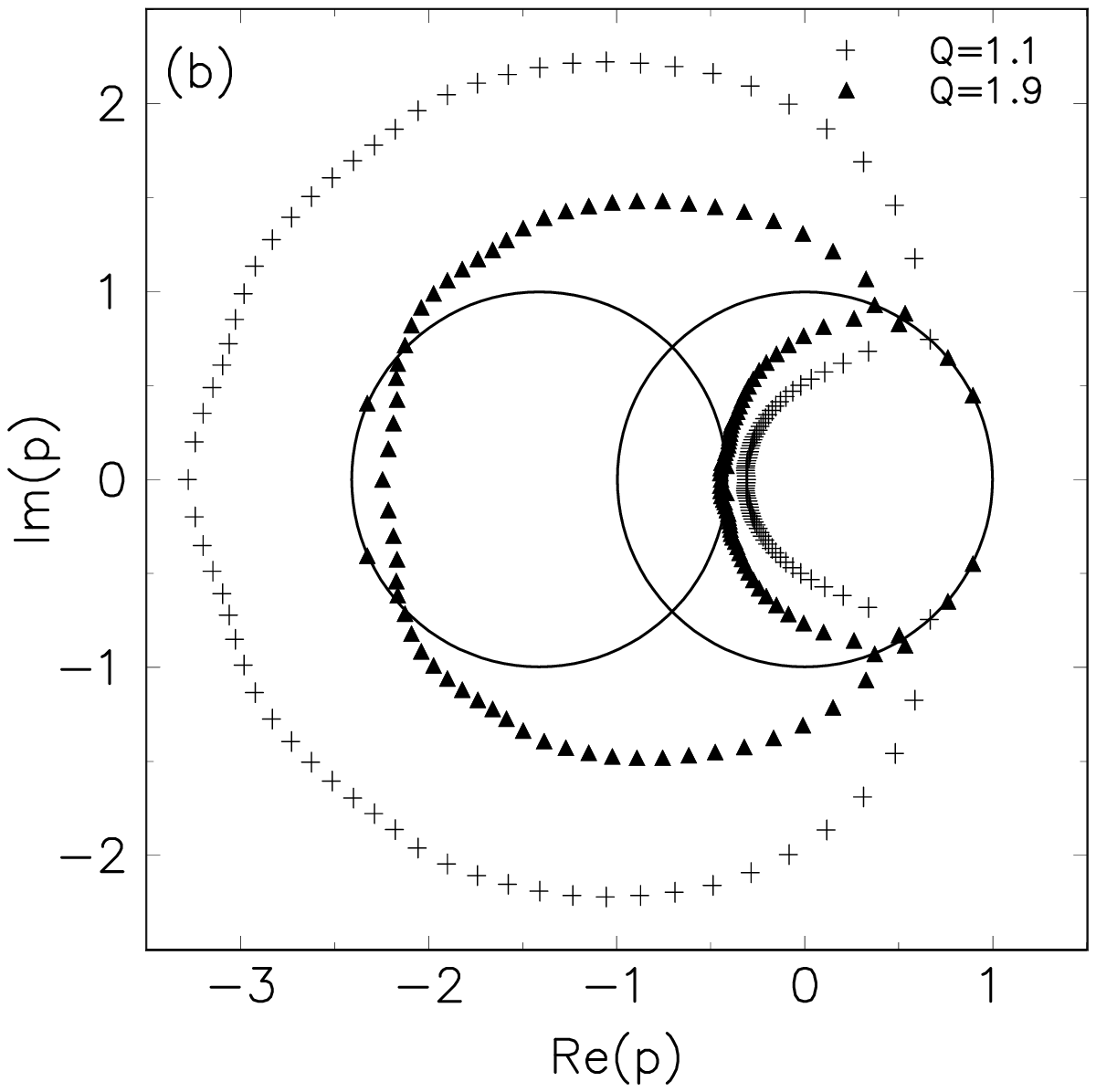}
\epsfbox{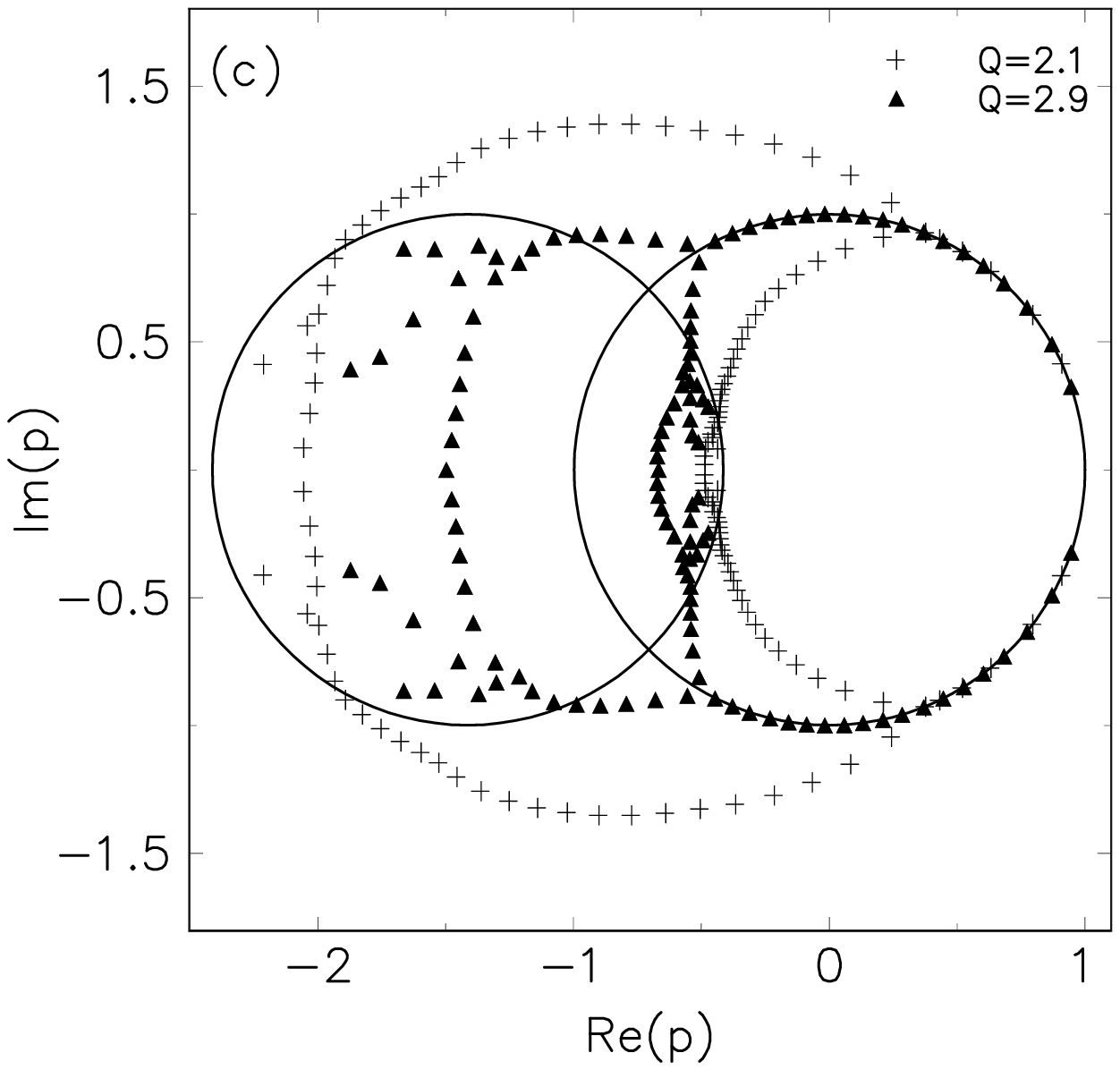}
\epsfbox{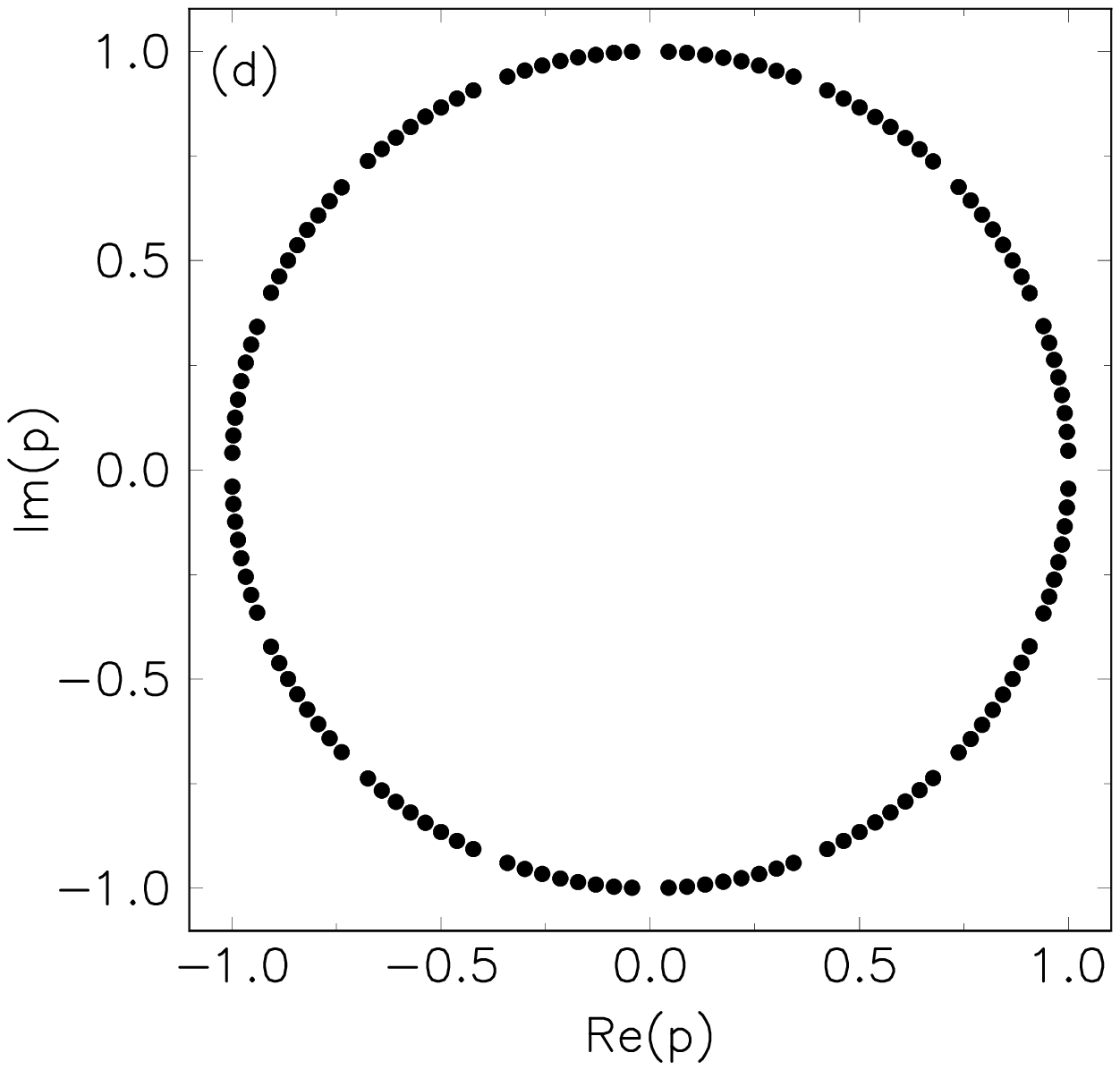}
\caption{Fisher zeros in the complex $p$ plane of 
$8\times8$ $Q$-state Potts models for
(a) $Q=0.1$ and 0.9, (b) 1.1 and 1.9, (c) 2.1 and 2.9, and (d) 9999.9.
In (b) and (c) two unit circles are the locus of the Fisher zeros
for the Ising model in the thermodynamic limit.}
\end{figure}

\begin{figure}
\epsfbox{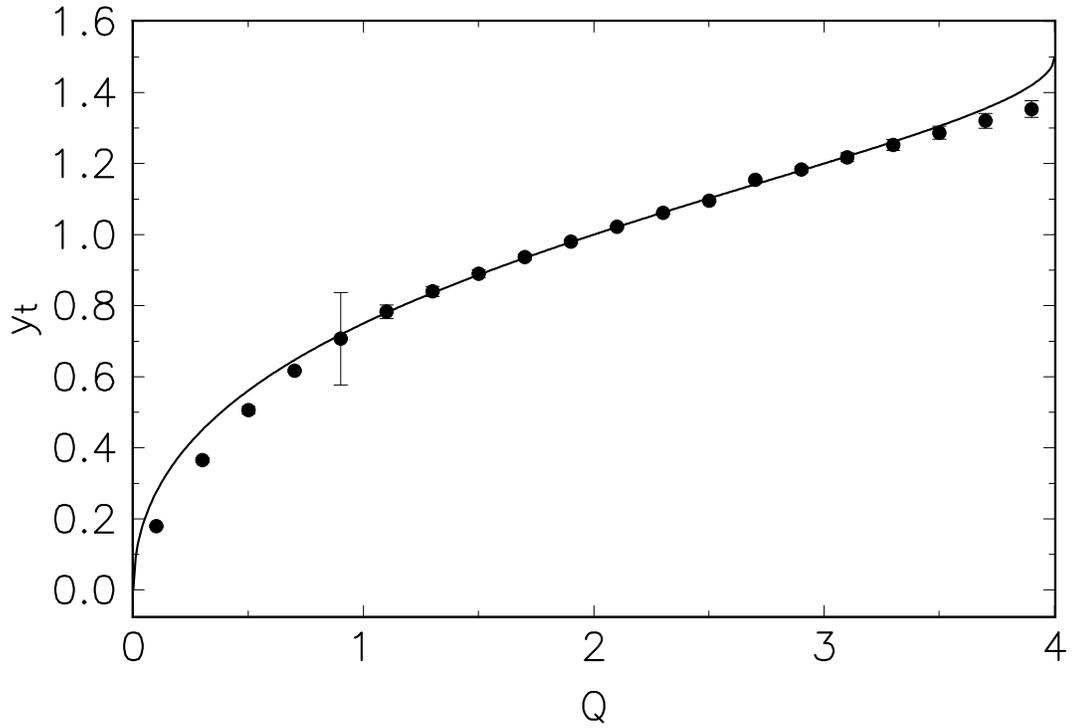}
\caption{The thermal exponent $y_t$ of the Potts ferromagnet
by the BST estimates (filled circles) and 
by the den Nijs formula (continuous curve).}
\end{figure}

\begin{figure}
\epsfbox{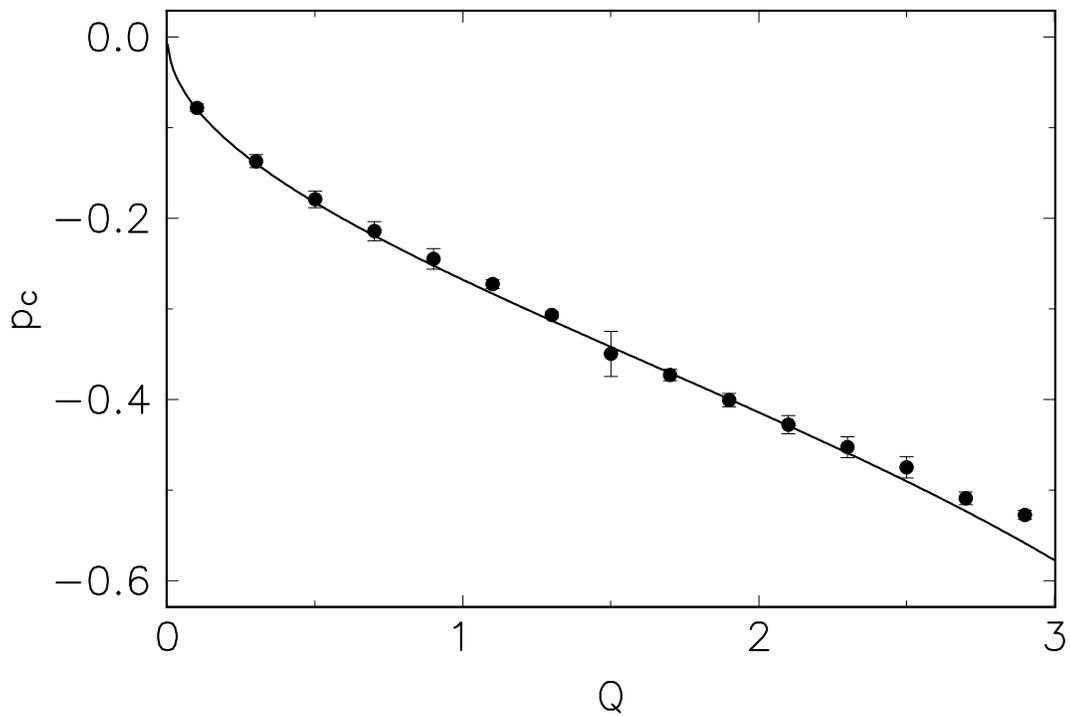}
\caption{The critical point of the Potts antiferromagnet 
by the BST estimates (filled circles) and by the Baxter conjecture
(continuous curve).}
\end{figure}

\begin{figure}
\epsfbox{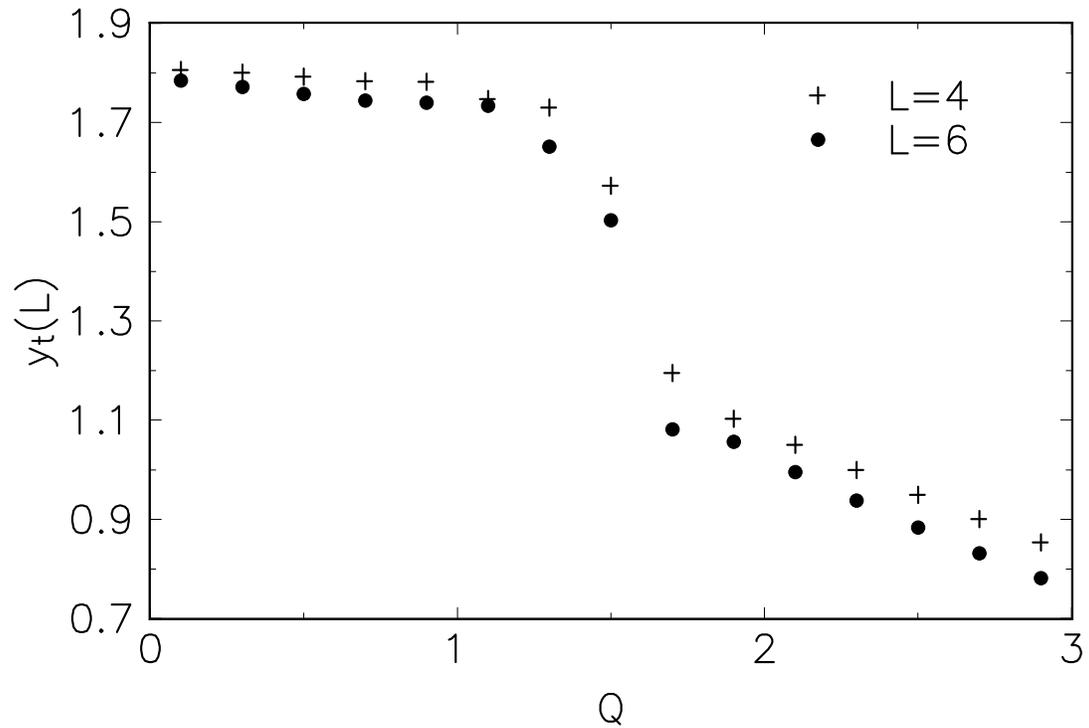}
\caption{The thermal exponent $y_t(L)$ of the Potts antiferromagnet
as a function of $Q$ for $L=4$ and $L=6$.
Note the large finite-size effect as $Q$ approaches 3.}
\end{figure}

\end{document}